\documentclass[9pt,conference]{IEEEtran}
\usepackage{waspaa25} % removes hyperlinks (required by IEEE Xplore)

\usepackage{bm} % for bold math symbols (incl. Greek letters) with \bm{}

\usepackage[table]{xcolor}
\usepackage{multicol}
\usepackage{multirow}
\usepackage{booktabs}
\usepackage{arydshln}
\usepackage{siunitx}
\usepackage{graphicx}
\usepackage{fontawesome5}

\usepackage{soul}% for underlines
\usepackage{colortbl} % for cell colors
\usepackage{changepage,threeparttable} % for wide tables

\title{OpenBEATs: A Fully Open-Source General-Purpose Audio Encoder}

\name{
Shikhar Bharadwaj$^{1}$,
Samuele Cornell$^{1}$,
Kwanghee Choi$^{1}$,
Satoru Fukayama$^{2}$, \\
\emph{Hye-jin Shim$^{1}$,
Soham Deshmukh$^{1}$,
Shinji Watanabe$^{1}$}
}

\address{
\cameraedit{$^{1}$Carnegie Mellon University, USA} \\
\cameraedit{
$^{2}$National Institute of Advanced Industrial Science and Technology (AIST), Japan }\\
\cameraedit{\texttt{sbharad2@andrew.cmu.edu}}
}

\begin{document}
\newcommand{\refalg}[1]{Algorithm \ref{#1}}
\newcommand{\refeqn}[1]{Equation \ref{#1}}
\newcommand{\reffig}[1]{Figure \ref{#1}}
\newcommand{\reftbl}[1]{Table \ref{#1}}
\newcommand{\refsec}[1]{Section \ref{#1}}
%\newcommand{\method}[1]{\mbox{\textsc{#1}}}

% Domain icons
\newcommand{\musicIcon}{\faMusic}
\newcommand{\soundIcon}{\faVolumeDown}
\newcommand{\bioIcon}{\faDog}
\newcommand{\ytIcon}{\faYoutube}

\newcommand{\bestnum}[1]{{\bfseries\num{#1}}}

\newcommand{\reminder}[1]{\textcolor{red}{[[ #1 ]]}\typeout{#1}}
\newcommand{\reminderR}[1]{\textcolor{gray}{[[ #1 ]]}\typeout{#1}}

\newcommand{\add}[1]{\textcolor{red}{#1}\typeout{#1}}
\newcommand{\remove}[1]{\sout{#1}\typeout{#1}}
\newcommand{\blue}[1]{\textcolor{blue}{#1}}
\newcommand{\cameraedit}[1]{#1}

\newcommand{\method}{OpenBEATs~}
\newcommand{\benchmark}{AudioVerse~}

\newcommand{\problem}{DD}
\newcommand{\problemfull}{Document Dating}

\newcommand{\mc}[1]{\mathcal{#1}}
\newcommand{\bmm}[1]{\bm{\mathcal{#1}}}
\newcommand{\real}[1]{\mathbb{R}^{#1}}

\newcommand{\tensor}{\mathcal{X}}
\newcommand{\Real}{\mathbb{R}}

\newcommand{\tuples}{\mathbb{T}}

\newcommand{\argmax}{arg\,max}

\newcommand\norm[1]{\left\lVert#1\right\rVert}

\newcommand{\note}[1]{\textcolor{blue}{#1}}

\newcommand*{\Scale}[2][4]{\scalebox{#1}{$#2$}}%
\newcommand*{\Resize}[2]{\resizebox{#1}{!}{$#2$}}%
\definecolor{officegreen}{rgb}{0.0, 0.5, 0.0}

%%% Tensor
%\DeclareMathAlphabet\ten{OMS}{cmsy}{b}{n} %%usage: \mathbfcal{W}
%% Matrix
\def\mat#1{\mbox{\bf #1}}%% usage: \mat{W}.

\maketitle

\begin{abstract}

Masked token prediction has emerged as a powerful pre-training objective across language, vision, and speech, offering the potential to unify these diverse modalities through a single pre-training task. 
\cameraedit{However, its application for general audio understanding remains underexplored, with BEATs being the only notable example.
BEATs has seen limited modifications due to the absence of open-source pre-training code.}
Furthermore, BEATs was trained only on AudioSet, restricting its broader downstream applicability.
To address these gaps, we present OpenBEATs, an open-source framework that extends BEATs via multi-domain audio pre-training.
We conduct comprehensive evaluations across six types of tasks, twenty five datasets, and three audio domains, including audio reasoning tasks such as audio question answering, entailment, and captioning.
OpenBEATs achieves state-of-the-art performance on six bioacoustics datasets, two environmental sound datasets and five reasoning datasets, performing better than models exceeding a billion parameters at one-fourth their parameter size.
These results demonstrate the effectiveness of multi-domain datasets and masked token prediction task to learn general-purpose audio representations.
To promote further research and reproducibility, we release all pre-training and evaluation code, pretrained and fine-tuned checkpoints, and training logs \footnote{\texttt{https://github.com/Shikhar-S/OpenBEATs}}.

\end{abstract}

\section{Introduction}
\label{sec:intro}

Self-supervised learning (SSL) has shown significant promise across a wide range of audio processing tasks. 
It allows models to learn general-purpose representations that transfer effectively to various downstream applications. 
Notable examples of SSL-based audio encoders (AEs) include BEATs \cite{beats}, SS-AST \cite{ssast}, MAE-AST \cite{maeast}, and Audio-MAE \cite{audiomae}. 
Among these, BEATs stands out due to the wide adoption of its pre-trained checkpoints, strong performance across DCASE challenges~\cite{cornell2024dcase,wu2023_t6a,beatsbiodcase} and the potential to unify vision, language and audio encoders through masked token prediction based learning.
However, the pre-training pipeline for BEATs remains closed-source, limiting such broader research impact.
In addition, BEATs uses a multistage pre-training setup, combining teacher-student distillation and masked audio modeling, making the pre-training more complex than other AEs.
In the speech community, open-sourcing models and their training pipelines has been a crucial practice \cite{openbestrq,owsm,wav2vec,hubert} that leads to enhanced reproducibility and accessibility.
Motivated by these gaps, this work aims to completely open-source the BEATs pre-training using the ESPnet toolkit \cite{espnet,versa} to foster further advancements.

Despite advances in SSL-based audio modeling, the training and evaluation of audio encoders remain fragmented across distinct domains -- environmental sound, bioacoustics and music.
As a result, state-of-the-art (SOTA) performance is typically achieved by domain-specific models.
For instance, BEATs excels in diverse environmental sound benchmarks, GPM-BT\cite{gpm-bt} leads in bioacoustics tasks, and MERT \cite{li2024mert} sets the bar for music-related tasks. 
In contrast, the speech community has demonstrated that multitask and multilingual training \cite{wavlm, usm, xeus} produces more generalizable and transferable representations.
Similar trends are evident in Audio-Language Models (ALMs), which unify diverse audio tasks under a shared sequence-to-sequence framework.
The performance of ALMs \cite{pengi,salmonn,gama} is often strongly correlated with the quality of the underlying audio encoder \cite{mellow}, underscoring the need for robust, general-purpose encoders. 
Yet, the current literature lacks both a universal audio encoder and a standardized cross-domain benchmark for evaluating generalization.
To address these limitations, we present a unified encoder trained on diverse audio domains.
Our results verify that multi-domain pre-training improves cross-domain generalization, achieving SOTA performance on multiple datasets.

Evaluation of audio encoders has largely been limited to classification tasks on a narrow set of general-purpose datasets, such as AudioSet \cite{audioset} and ESC-50 \cite{esc50}. 
Broader evaluations -- particularly in specialized domains like bioacoustics -- remain underexplored or are deferred to downstream users. 
However, with the advent of ALMs, audio encoders are increasingly being used for more complex semantic tasks, such as audio captioning, audio question answering, and audio reasoning, spanning diverse domains. 
In this setup, the audio encoder is expected to contain both audio information and necessary semantic information to steer the language model. 
This shift highlights the need for comprehensive evaluation of audio representations not only on traditional classification tasks, but also on open-ended and semantically rich tasks across bioacoustics, environmental sound, and music domains.
To support this, we benchmark multiple models and offer a streamlined interface, via a single script, for testing new encoders using ESPnet.

\begin{figure*}[t]
    \centering
    \includegraphics[width=0.9\textwidth]{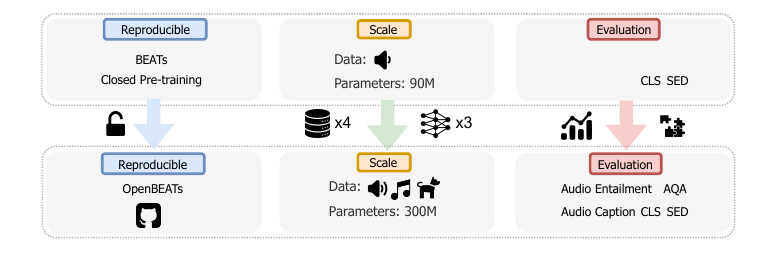}
    \vspace{-1.5em}
    \caption{Our contributions in this work are three folds: 1) Open sourcing the pre-training and evaluation code for BEATs. 2) Scaling the training data and model to handle multiple sound domains like music (\musicIcon), environmental sound (\soundIcon) and bioacoustics (\bioIcon). 3) Fine‑grained multi-domain evaluations that go beyond standard tasks, probing reasoning ability via audio entailment and question-answering.}
    \label{fig:main}
\end{figure*}

In this work, we address the aforementioned challenges, by releasing an open-source and reproducible training pipeline for BEATs, scaling it across diverse domains, and proposing a unified benchmark for comprehensive ALM-focused evaluation.
The main contributions, also depicted in \cref{fig:main}, are:
\begin{itemize}
    \item We open-source the training and evaluation framework for BEATs, a SOTA SSL audio encoder. 

    \item We scale the BEATs model to achieve general-purpose audio representations that can be used for a variety of downstream tasks. 
    Specifically, we increase the model size to 300M parameters and expand the pre-training data to 20k hours by adding environmental sound datasets and incorporating bioacoustics and music domains.
    Our model achieves strong cross-domain transfer and is comparable to models with over a billion parameters \cite{dasheng}.

    \item We introduce an open-source comprehensive evaluation suite that spans multiple domains, including bioacoustics, music and environmental sound.
    Our benchmark is designed to streamline the evaluation of audio encoders and include tasks such as audio entailment, question answering, and captioning.
\end{itemize}

\section{Background}
\label{sec:background}

\subsection{BEATs Architecture}
BEATs (Bidirectional Encoder representation from Audio Transformers) \cite{beats} is based on a Vision Transformer \cite{vit} architecture adapted for audio representation learning.
\cameraedit{The model processes 16 kHz audio by converting it into 128-dimensional mel spectrograms, which are segmented into 16×16 patches and then fed to a transformer encoder.}
This patching mechanism enables BEATs to capture both local and global spectral features efficiently. 
The backbone of the model consists of stacked transformer encoder blocks, each employing multi-head self-attention to learn contextual representations of audio signals.

\subsection{BEATs Pre-training}
BEATs undergoes two-stage SSL, involving encoder training and tokenizer training.
These stages iteratively refine both the feature representations and the discrete tokenization of audio signals, improving the model’s ability to learn structured embeddings, with the process repeated up to three times.

\subsubsection{Encoder Training}
The pre-training of BEATs encoder follows a Masked Language Modeling (MLM) paradigm applied to audio.
In the first iteration, masked spectrogram patches are tokenized using a randomly initialized tokenizer \cite{bestrq}.
From the second iteration onward, the model employs a trained tokenizer, leading to progressively refined learned representations.
Given an input sequence of $d$ dimensional Mel-spectrogram patches with length $N$, $\mathbf{X} = \{\mathbf{x}_n \in \mathbb{R}^d\}_{n=1}^N$, a subset of patches is randomly masked, producing a sequence of unmasked patches \( \tilde{\mathbf{X}} \).
The encoder produces a sequence of $h$-dimensional embeddings \( f^e(\tilde{\mathbf{X}}) = \{\mathbf{o}_n \in \mathbb{R}^{h}\}_{n=1}^N\), and it is fed through the label predictor module $f^e_\text{pred}$.
Then, \(f^e_\text{pred} \circ f^e\) is learned to predict the masked tokens by minimizing the cross-entropy loss between the predicted tokens and the ground truth token indices given the unmasked patches:
\begin{equation}
\mathcal{L}_\text{MLM} = - \sum_{n \in M} \log p(z_n | \tilde{X}),
\label{eq:mlm_loss}
\end{equation}
where \( M \) is the set of masked positions and \( z_n \) represents the ground truth token for position \( n \) yielded from the tokenizer. Therefore, $\log p(z_n | \tilde{X})$ stands for the predicted probability distribution of \( p \) over possible tokens at position \( n \), given the masked input \( \tilde{\mathbf{X}} \).
At the implementation level, besides tracking the loss in \cref{eq:mlm_loss}, metrics like vocabulary coverage and masked accuracy are also tracked in our open-source implementation.

\begin{table*}[t]
  \caption{Architecture, data, and training configuration for BEATs and \method variants.}
  \label{tbl:model_config}
  \centering
  \scriptsize
  \resizebox{\linewidth}{!}{%
  \begin{tabular}{lcccccccccccc}
    \toprule
    \textbf{Model} & \textbf{Data (hr)} & \textbf{Param} & \textbf{Hidden size} & \textbf{FFN size} & \textbf{Layers} & \textbf{\# Heads} & \textbf{\# Codebook} & \textbf{Batch (sec)} & \textbf{Updates} & \textbf{Warmup} & \textbf{LR} & \textbf{Opt.} \\
    \midrule
    BEATs & 5.8k & 90M & 768 & 3072 & 12 & 12 & 1024 & 4.6k & 400k & 32k & 5e-4 & AdamW \\
    \method-Base & 20k & 90M & 768 & 3072 & 12 & 12 & 1024 & 10.7k & 400k & 40k & 5e-4 & AdamW \\
    \method-Large & 20k & 300M & 1024 & 4096 & 24 & 16 & 1024 & 10.7k & 400k & 40k & 1e-4 & AdamW \\
    \bottomrule
  \end{tabular}
  }
\end{table*}

\subsubsection{Tokenizer Training}
BEATs adopts a teacher-student framework for tokenizer training.
The teacher model corresponds to the encoder above from the previous iteration, while the student model is the tokenizer being trained in the current iteration.

The tokenizer encodes input \( \mathbf{X}\) into embedding sequence \( f^t(\mathbf{X}) = \mathbf{E} = \{\mathbf{e}_n \in \mathbb{R}^{p}\}_{n=1}^N\), which is quantized into \( q(\mathbf{E}) = \{\mathbf{v}_{z_n} \in \mathbb{R}^{p}\}_{n=1}^N \).
Here, the token \(z_n\) is the nearest neighbor index in the tokenizer codebook \( \mathbf{V} = \{\mathbf{v}_1, \mathbf{v}_2, \cdots, \mathbf{v}_K \}\) of size \(K\):
\begin{equation}
    z_n = \arg\min_{t} \left| \bar{\mathbf{v}}_t - \bar{\mathbf{e}}_n \right|_2^2,
\label{eq:znn}
\end{equation}
where \( \bar{\mathbf{v}}_t\) and \(\bar{\mathbf{e}}_n\) is the L2-normalized \({\mathbf{v}}_t\) and \( {\mathbf{e}}_n \).

Given the quantized embedding sequence \( q(\mathbf{E}) \), the tokenizer learns to predict embeddings \( f^t_\text{pred}(q(\mathbf{E})) = \{\mathbf{\hat{o}}_n \in \mathbb{R}^p \}_{n=1}^N \) to align with the teacher embeddings through the knowledge distillation loss $\mathcal{L}_\text{KD}$:
\begin{equation}
\label{eq:vecquant}
\begin{aligned}
    \mathcal{L}_\text{KD} &= -\sum_n \cos( \mathbf{\hat{o}}_n, \mathbf{o}_n ) +  \left| \operatorname{sg}[\bar{\mathbf{e}}_n] - \bar{\mathbf{v}}_{{z}_n} \right|_2^2 + \left| \bar{\mathbf{e}}_n - \operatorname{sg}[{\bar{\mathbf{v}}_{{z}_n}}] \right|_2^2,
\end{aligned}
\end{equation}

where
\( \operatorname{sg}[\cdot] \) is the stop gradient operator.
\Cref{eq:vecquant} has three terms: a cosine similarity loss encouraging alignment between the teacher and tokenizer, and two L2 loss terms improving stability and consistency of codebook representations.
At the implementation level, the gradients of the quantized embedding sequence $q(\mathbf{E})$ are copied to the embedding sequence $\mathbf{E}$ to solve the non-differentiable nature of vector quantization \cite{vecquant}, and the codebook \(\mathbf{V}\) is initialized using K-Means and optimized through exponential moving average.

\section{\method}
\label{sec:expdetails}

In this work, we extend the pre-training data beyond AudioSet~\cite{audioset} by incorporating additional large-scale public datasets such as FreeSound, FMA \cite{fma}, and iNaturalist~\cite{inat}, which span diverse domains including music, environmental sound, and bioacoustics.
In prior literature \cite{biolingual, gpm-bt} bioacoustics datasets have been used in isolation and not combined with music.
However, to the best our knowledge, this work is the first to systematically unify these diverse domains and study the cross-domain impact from such multi-domain training via extensive evaluations.
In addition, our model is trained with variable-length input sequences, making it more adaptable to the natural variability of real-wold audio data.
Together, these changes enable our encoder to learn robust, general purpose representations for a wide range of tasks and domains. 
A detailed list of our pre-training datasets is provided in~\cref{tab:pretraining_data}.
\cameraedit{We exclude speech corpora, as speech requires fine-grained temporal resolution to model phoneme durations, which differs from the coarser time resolution of 175 ms employed in a BEATs patch.}

We first train a 90M parameter \method model (\texttt{base}) on this expanded dataset and observe saturation in downstream task performance. 
To address this, we scale up the model architecture and train a 300M parameter variant (\texttt{large}), following the \texttt{base}/\texttt{large} design pattern established by HuBERT~\cite{hubert}.
\Cref{tbl:model_config} lists the details of our architecture and pre-training hyper-parameters.
To support reproducibility and future research, we open-source the full training pipeline, including the pre-training code, data preprocessing scripts, and training logs for all our models.

\vspace{-2mm}
\begin{table}[h]
    \centering
    \small
    \caption{Overview of pre-training datasets covering music (\musicIcon), bioacoustics (\bioIcon) and environmental sound events (\soundIcon). *We use data filtered using the WavCaps \cite{wavcaps} processing pipeline for these sources.}
    \begin{tabular}{@{}l c c c@{}}
    \toprule
    Dataset & \multicolumn{1}{c}{Hours} & \multicolumn{1}{c}{Instances} & Domain \\
    \midrule
    FMA \cite{fma} & 7.7k & 2.8M & \musicIcon \\
    AudioSet \cite{audioset} & 5.6k & 2.0M & \soundIcon \\
    FreeSound\textsuperscript{*} & 4.1k & 1.7M & \soundIcon \\
    BBC Sound Effects\textsuperscript{*} & 1.0k & 0.4M & \soundIcon \\
    iNat Sounds \cite{inat} & $0.8$k & 0.3M & \bioIcon \\
    Others \cite{cochlscene,epickitchen} & $0.3$k & $0.1$M & \soundIcon \\
    \midrule
    \textbf{Total} & \textbf{20k} & \textbf{7.3M} & \\
    \bottomrule
    \end{tabular}
    
    \label{tab:pretraining_data}
\end{table}

% \footnotetext[2]{\texttt{https://freesound.org/}}
% \footnotetext[3]{\texttt{https://sound-effects.bbcrewind.co.uk/}}

\vspace{-5mm}
\begin{table}[!htp]
\centering
\caption{
Evaluation tasks. Recipes for all of these tasks are also open-sourced as part of this work.
LP = Linear Probing; FT = Full Fine-tuning; DFT = Decoder Fine-Tuning; CLS = Classification; SED = Sound Event Detection.
We follow prior works for evaluation protocols where available (eg. BEANS).
Higher values are better for all metrics. See details in \cref{sec:setupdetails}.
}
\label{tbl:evaltasks}
\resizebox{\columnwidth}{!}{%
\begin{tabular}{p{2.5cm} p{1.2cm} p{1.3cm} p{0.7cm} p{1.1cm} p{1.4cm}}
\toprule
\textbf{Dataset} & \textbf{Abbr.} & \textbf{Task Type} & \textbf{Eval} & \textbf{Metric} & \textbf{Reference} \\
\midrule
\multicolumn{6}{l}{\textit{Environmental Sound}} \\
\midrule
DESED & DSD & SED & LP & Score & \cite{desed} \\
UrbanSound8K & US8K & CLS & LP & Score & \cite{us8k} \\
FSD-50K & F50K & SED & LP/FT & Score/mAP & \cite{fsd50k} \\
ESC-50 & ESC & CLS & LP/FT & Score/Acc & \cite{esc50} \\
FSD2018-Kaggle & FKgl & SED & LP & Score & \cite{fsdkgl} \\
Clotho (Retrieval) & Clotho & Retrieval & LP & Score & \cite{clotho} \\
AudioSet-2M & AS-2M & SED & FT & mAP & \cite{audioset} \\
AudioSet-20K & AS-20K & SED & FT & mAP & \cite{audioset} \\
\midrule
\multicolumn{6}{l}{\textit{Bioacoustics}} \\
\midrule
BEANS (10 tasks) & BEANS & CLS/SED & FT & Acc/mAP & \cite{beans} \\
\midrule
\multicolumn{6}{l}{\textit{Reasoning}} \\
\midrule
Audio Entailment & Entailment & Entailment & FT & Acc & \cite{audioentailment} \\
AudioQA & AQA & Question Answering & FT & Acc & \cite{audioqa} \\
Clotho (Captioning) & Clotho AAC & Captioning & DFT & CIDEr\cite{cider} & \cite{clotho} \\
\midrule
\multicolumn{6}{l}{\textit{Music}} \\
\midrule
GTZAN & GTZAN & CLS & FT & Acc & \cite{gtzan, gtzan_flaw} \\
NSynth-Instr. & NSynth-I & CLS & FT & Acc & \cite{nsynth} \\
NSynth-Pitch & NSynth-P & CLS & FT & Acc & \cite{nsynth} \\
\bottomrule
\end{tabular}%
}
\end{table}

\vspace{-3mm}
\section{Evaluation Setup}  
\label{sec:evals}
\begin{table}[t]
\centering
\caption{Linear probing results on sound tasks from X-ARES benchmark \cite{icme}. \texttt{B} and \texttt{L} denotes \texttt{Base} and \texttt{Large} models. \method achieves best performance on DESED and UrbanSound-8K datasets, and performs competitively with other models of similar scale across the remaining datasets.}
\label{tab:linprobe}
\sisetup{
    round-mode=places,
    round-precision=2,
    table-format=1.3,
    table-number-alignment=center,
    detect-weight=true,
    detect-family=true,
}

\setlength{\tabcolsep}{1pt}
\begin{tabular}{
    l
    c
    *{6}{S[table-format=1.2]}
    S[table-format=1.3,table-column-width=1.5em]
}
\toprule
\textbf{Model} & \textbf{Param} & \textbf{DSD} & \textbf{US8k} & \textbf{F50k} & \textbf{ESC} & \textbf{FKgl} & \textbf{Clotho} & \textbf{Avg} \\
\midrule
Dasheng \texttt{B} \cite{dasheng}         & 90M & 0.532 & 0.835 & 0.408 & 0.869 & 0.557 & 0.033 & 0.539 \\
Data2Vec \cite{baevski2022data2vec}           & 90M & 0.137 & 0.443 & 0.084 & 0.249 & 0.196 & 0.006 & 0.186 \\
Whisper \cite{radford2023robust}           & 74M & 0.125 & 0.719 & 0.262 & 0.614 & 0.478 & 0.029 & 0.371 \\
EAT \texttt{B} \cite{chen2024eat}              & 88M & 0.356 & 0.792 & 0.359 & 0.658 & 0.383 & 0.042 & 0.432 \\
BEATs (iter3) \cite{beats}      & 90M & 0.560 & 0.853 & 0.217 & 0.835 & 0.545 & 0.042 & 0.509 \\
EAT \texttt{L}  \cite{chen2024eat}            & 309M & 0.439 & 0.819 & 0.435 & 0.732 & 0.577 & \bestnum{0.057} & 0.510 \\
Dasheng-0.6B \cite{dasheng}      & 600M & 0.541 & 0.848 & 0.445    & 0.883 & 0.576 & 0.037 & 0.555 \\
Dasheng-1.2B \cite{dasheng}     & 1.2B & 0.563 & 0.846 & \bestnum{0.455}    & \bestnum{0.891} & \bestnum{0.627} & 0.036 & 0.570 \\
\midrule
\method \texttt{L}          & 300M &\bestnum{0.565} & \bestnum{0.866} & 0.434 & 0.859 & 0.544 & 0.046 & 0.552 \\
\bottomrule
\end{tabular}
\end{table}

\begin{table}[!htp]
\centering
\caption{\method achieves best performance on environmental sound detection tasks. $^\dag$ denotes values reported by BEATs, all other values are from our open-source implementation. Please note that our copy of AudioSet (AS) downloaded from YouTube differs from BEATs. See \cref{sec:results} for details.}
\resizebox{\linewidth}{!}{%
\begin{tabular}{lccccc}
\toprule
\textbf{Model} & \textbf{Param} & \textbf{ESC-50} & \textbf{FSD-50K} & \textbf{AS-2M} & \textbf{AS-20K} \\
& & acc & mAP & mAP & mAP \\
\midrule
Audio-MAE \cite{audiomae}  & 86M & 94.1$^\dag$ & - & 47.3$^\dag$ & 37.1$^\dag$ \\
Audio-MAE & 304M & - & - & 47.4$^\dag$ & 37.6$^\dag$ \\
\midrule
BEATs iter 1 & 90M & 93.0$^\dag$ & 52.3 & 47.9$^\dag$ & 36.0$^\dag$ \\
BEATs iter 2 & 90M & 95.1$^\dag$ & 55.4 & 48.1$^\dag$ & 36.0$^\dag$ \\
BEATs iter 3 & 90M & 95.6$^\dag$ / 94.8 & 56.2 & 48.0$^\dag$ / 41.6 & 38.3$^\dag$ / 34.1 \\
\midrule
\method~iter 1 & 90M & 93.9 & 54.4 & 39.4 & 31.5 \\
\method~iter 2 & 90M & 94.8 & 55.8 & 41.1 & 31.5 \\
\method~iter 3 & 90M & 95.0 & 55.5 & 40.8 & 32.1 \\
\midrule
\method~iter 1 & 300M & 95.3 & 56.7 & 41.5 & 32.7 \\
\method~iter 2 & 300M & 95.7 & \textbf{57.5} & 42.2 & 32.4 \\
\method~iter 3 & 300M & \textbf{95.8} & 56.8 & 42.1 & 33.1 \\
\bottomrule
\end{tabular}%
}
\label{tbl:envsound}
\end{table}
\vspace{-2mm}

\renewcommand{\arraystretch}{0.83}
\begin{table*}[!htp]
\centering
\cameraedit{\caption{Bioacoustics: Full-finetuning evaluations on BEANS \cite{beans}. \method achieves SOTA results performing competitively with models pre-trained with \textcolor{gray}{audio-text supervision}
\cite{laion-clap,biolingual}, in-domain SSL \cite{gpm-bt,aves}, and SSL models with more parameters \cite{dasheng}.}}

\scriptsize
\begin{tabular}{lcc*{5}{>{\columncolor{blue!5}}c}*{5}{>{\columncolor{green!5}}c}}
\toprule
\multirow{2}{*}{\textbf{Model}} & \multirow{2}{*}{\textbf{Pre-training}}  & \multirow{2}{*}{\textbf{\#Params}} & \multicolumn{5}{c}{\textbf{Sound Event Classification (acc)}} & \multicolumn{5}{c}{\textbf{Sound Event Detection (mAP)}} \\
\cmidrule(lr){4-8} \cmidrule(lr){9-13}
& \textbf{Data Instances} & & \textbf{Watkins} & \textbf{CBI} & \textbf{HumBugDB} & \textbf{Dogs} & \textbf{Bats} & \textbf{DCASE 21} & \textbf{Rfcx} & \textbf{Gibbons} & \textbf{Hiceas} & \textbf{Enabirds} \\
\midrule

\textbf{Audio-text CL} & & & & & & & & & & & & \\
LAION-CLAP \cite{laion-clap} & 633k pairs & \textcolor{gray}{31M} & \textcolor{gray}{89.1} & \textcolor{gray}{62.2} & \textcolor{gray}{82.0} & \textcolor{gray}{96.4} & \textcolor{gray}{75.3} & \textcolor{gray}{47.0} & \textcolor{gray}{14.5} & \textcolor{gray}{29.6} & \textcolor{gray}{62.7} & \textcolor{gray}{66.4} \\

BioLingual \cite{biolingual} & 1.1M pairs & \textcolor{gray}{31M}  & \textcolor{gray}{89.4} & \textcolor{gray}{74.4} & \textcolor{gray}{81.7} & \textcolor{gray}{97.1} & \textcolor{gray}{76.6} & \textcolor{gray}{47.5} & \textcolor{gray}{17.8} & \textcolor{gray}{37.6} & \textcolor{gray}{67.7} & \textcolor{gray}{68.8} \\

\midrule
\textbf{Bioacoustics SSL} & & & & & & & & & & & & \\
AVES \cite{aves} & 1.8M & 89M & 87.9 & 59.8 & 81.0 & 95.0 & 74.8 & 39.2 & 13.0 & 28.4 & 62.9 & 55.5 \\
GPM-BT \cite{gpm-bt} & 1.2M & 89M & \textbf{91.4} & 54.3 & 81.1 & 94.2 & 77.4 & 45.4 & 12.9 & 34.5 & 65.0 & \textbf{62.4} \\

\midrule
& & 90M & 77.3 & 63.1 & 69.9 & 95.7 & 75.5 & 45.6 & 11.7 & 41.8 & 43.8 & 43.9 \\
\textbf{Dasheng} \cite{dasheng} & 97M & 600M & 79.7 & 67.8 & 73.6 & 95.7 & 77.3 & 45.5 & 13.6 & 47.3 &  58.8 & 45.6 \\
& & 1.2B & 79.1 & 66.7 & 72.2 & \textbf{97.1} & 76.5 & 46.4 & 13.2 & 36.8 & 57.4 & 45.0 \\

\midrule
\textbf{BEATs} & & & & & & & & & & & & \\
iter 1 & & & 87.0 & 61.9 & 80.1 & \textbf{97.1} & 78.8 & 45.0 & 8.6 & 45.8 & 71.3 & 48.8 \\
iter 2 & 2M & 90M & 87.9 & 64.4 & 87.3 & 92.1 & 79.3 & 45.2 & 10.6 & 44.0 & \textbf{69.5} & 51.1 \\
iter 3 & & & 89.4 & 64.0 & 80.6 & 90.7 & 76.8 & 43.8 & 10.3 & 44.8 & 67.4 & 49.6 \\

\midrule
\midrule
\textbf{\method Base} & & & & & & & & & & & & \\
iter 1 & & & 85.0 & 63.6 & 84.2 & 95.0 & 78.4 & 44.5 & 10.1 & 45.6 & 68.2 & 46.2 \\
iter 2 & 7.4M & 90M & 86.4 & 64.9 & 87.3 & 96.4 & 77.3 & 46.2 & 13.2 & 47.3 & 68.2 & 49.0 \\
iter 3 & & & 87.0 & 64.5 & 84.8 & 95.7 & 76.9 & 46.4 & 12.4 & 42.3 & 68.1 & 49.0 \\
\midrule
\textbf{\method Large} & & & & & & & & & & & & \\
iter 1 & & & 88.8 & 67.8 & 86.6 & 95.0 & 79.4 & \textbf{49.5} & 13.0 & 45.3 & 69.1 & 52.1 \\
iter 2 & 7.4M & 300M & 88.8 & 68.8 & \textbf{87.7} & 95.7 & \textbf{79.7} & 47.6 & 13.0 & \textbf{50.2} & 68.3 & 53.4 \\
iter 3 &  & & 88.2 & \textbf{69.4} & 87.4 & 95.7 & 79.5 & 46.2 & \textbf{14.9} & 49.3 & 66.0 & 54.4 \\

\bottomrule
\label{tbl:bioacoustics}
\end{tabular}
\vspace{-2mm}
\end{table*}
\renewcommand{\arraystretch}{1.0}

\begin{table}[!htp]
\centering
\caption{Audio reasoning tasks. \method outperforms other encoders on entailment and AQA and performs competitively with BEATs on captioning.}
\scriptsize
\begin{tabular}{l cc cc c}
\toprule
\multirow{2}{*}{\textbf{Model}} & \multicolumn{2}{c}{\textbf{AQA (BERT / CLAP)}} & \multicolumn{2}{c}{\textbf{Entailment}} & \multirow{2}{*}{\textbf{Clotho}} \\
\cmidrule(lr){2-3} \cmidrule(lr){4-5}
& \textbf{Open} & \textbf{Yes-No} & \textbf{BERT} & \textbf{CLAP} & \textbf{AAC}\\
\midrule
Metrics & acc & acc & acc & acc & CIDEr \\
\midrule
BEATs & 10.5 / 9.7 & 56.7 / 57.5 & 73.0 & 71.5 & 39.0 \\
Dasheng-1.2B & 10.4 / 10.1 & 58.5 / 56.7 & \textbf{74.5} & 73.0 & \textbf{47.7} \\
\method Base & 12.2 / 11.1 & 59.4 / \textbf{59.4} & 73.2 & \textbf{75.2} & 35.3 \\
\method Large & \textbf{12.7} / \textbf{11.5} & \textbf{59.9} / 58.7 & 73.2 & 74.2 & 37.5 \\
\bottomrule
\end{tabular}
\label{tbl:reasoning}
\end{table}

\begin{table}[!htp]
\centering
\cameraedit{\caption{\method improves performance on two of the three music tasks.}}
\label{tbl:music}
\scriptsize
\begin{tabular}{lccc}
\toprule
\textbf{Model} & \textbf{GTZAN Genre} & \textbf{NSynth-I} & \textbf{NSynth-P} \\
\midrule
BEATs & 87.1 & 79.9 & \textbf{93.1} \\
Dasheng Base & 87.1 & 80.9 & 90.9  \\
\method Base & 88.1 & 79.5 & 92.5  \\
\method Large & \textbf{89.1} & \textbf{81.7} & 92.7 \\
\bottomrule
\end{tabular}
\end{table}

\vspace{-2mm}
\subsection{Baselines}

\noindent \textbf{Dasheng}:
Dasheng \cite{dasheng} surpasses the billion-parameter mark and is pre-trained on over 272k hours of data, covering multi-domain and speech-rich sources from YouTube.
In contrast, our setup offers a more controlled comparison by deliberately restricting to audio-only datasets.
Dasheng adopts a mean squared error loss, whereas \method relies on masked modeling with discrete tokens. 

\noindent \textbf{Speech models}:
To gauge the transferability of speech-only pre-training, we compare with Data2Vec\cite{baevski2022data2vec} and Whisper-small\cite{radford2023robust}.

\noindent \textbf{Bioacoustics models}:
We evaluate domain-specific pre-training via dedicated bioacoustics sound encoders like AVES \cite{aves}, BioLingual \cite{biolingual}, to test the claim that multi-domain SSL can outperform models trained on specific datasets.
We also report GPM-BT \cite{gpm-bt} which has shown promising results on bioacoustics.
While GPM-BT and AVES are SSL models, BioLingual is trained with text supervision following a CLAP-like training methodology.

\noindent \textbf{AudioMAE}:
AudioMAE employs a pure mask-and-reconstruct objective.
This is an alternate SSL methodology to BEATs as it does not use any tokenization.
Numbers are reported on standard audio classification tasks like ESC-50 and AudioSet.

\noindent \textbf{Efficient Audio Transformer (EAT)}: 
EAT \cite{chen2024eat} is another recently proposed SOTA audio encoder trained on AudioSet \cite{audioset}.

\subsection{Evaluation Tasks}
\label{sec:setupdetails}
\Cref{tbl:evaltasks} lists all evaluation tasks. 
Linear probing measures the effectiveness of pretrained representations without any fine-tuning induced variations \cite{noss}.
We benchmark linear probes on sound tasks using the X-ARES toolkit \footnote{\texttt{https://github.com/jimbozhang/xares/tree/main}}.
This toolkit converts all individual metrics to be in the range of 0-1, with higher values being better.
The sound reasoning tasks evaluate the utility of audio representations even before integrating with frozen language models to build ALMs.
We consider two setups: (1) a multi-modally trained text encoder, like CLAP\cite{laion-clap}, and (2) an independently trained text encoder, like BERT.
By comparing these approaches, we can understand how an audio encoder will behave with audio-informed text embeddings.
Therefore, our benchmark bridges the gap between low-level feature extraction and high-level cognitive reasoning, offering a structured framework to improve audio encoders for ALMs. 
For audio captioning, our setup consists of the frozen audio encoder and a trainable transformer decoder, initialized randomly \cite{dcase24aac_top}.

\section{Results and Discussion}
\label{sec:results}

\noindent \textbf{Linear probing on X-ARES:}
As summarized in \cref{tab:linprobe}, \method \texttt{Large} achieves the best linear-probe accuracy on DESED and US8k, and remains within a few points of the 1.2 billion parameter Dasheng on FSD50K and ESC-50.
Notably, it does so with four times fewer parameters than Dasheng-1.2B and a considerably smaller pre-training data budget. 
These results show that token prediction based modeling approach scales more efficiently, delivering competitive performance with less data.

\noindent \textbf{Environmental Sound:}
\method consistently delivers strong performance on the canonical ESC-50 and FSD-50K benchmarks in \cref{tbl:envsound}.
The \texttt{Large} model attains 95.8 \% accuracy on ESC-50 and 57.5 mAP on FSD-50K, confirming that parameter scaling pays off after multi-domain data saturates the 90 M scale.
Performance gaps in AudioSet originate primarily from differences in the evaluation dataset, since YouTube is not a static source and our evaluation set differs by 5\% from the one reported in BEATs.
We verified the correctness of our training and inference code by replicating the numbers using the public BEATs inference codebase.
We release both codebases and the YouTube IDs for our copy of the AudioSet publicly.
The improved performance from \method \texttt{Large} on AS-2M shows the benefits of multi-domain pre-training.
These results demonstrate that the pre-training technique scales gracefully.

\noindent \textbf{Bioacoustics:}
\Cref{tbl:bioacoustics} shows that the \method attains SOTA results on 6 of the 10 BEANS datasets among all SSL models. 
Even against billion-parameter Dasheng variants, \method \texttt{Large} remains better confirming that performance scales favourably with masked token prediction based modeling applied to multi-domain datasets.
Crucially, \method also outperforms the text-supervised audio–text contrastive baselines like LAION-CLAP and BioLingual, demonstrating that scaled-up SSL can surpass models that have access to paired textual information.
These findings show that multi-domain pre-training yields general-purpose representations.

\noindent \textbf{Sound Reasoning and Music Tasks:}
\method achieves best accuracies on both AQA and entailment tasks, underscoring the benefits of its larger capacity and broader pre-training corpus (\cref{tbl:reasoning}). 
Consistently, we observe that scaling the model size and expanding pre-training data yields systematic gains for both Dasheng and OpenBEATs, confirming that representation quality improves with greater model and data scale. 
Nevertheless, OpenBEATs lags behind on audio-captioning.
One possible reason is that we applied the recipe tuned for BEATs by \cite{dcase24aac_top} to other models without sufficient hyper-parameter exploration.
\method also improves performance on music tasks (\cref{tbl:music}), showing that multiple domains in pre-training data produce general purpose representations.

\section{Conclusion}
\label{sec:conclusion}

In this work, we open-source the complete pre-training pipeline for BEATs and demonstrate its effectiveness through strong performance across diverse benchmarks. Our comprehensive evaluation spans multiple domains and tasks, showcasing the robustness and versatility of the learned representations. 
We find that multi-domain pre-training enables the model to generalize effectively, achieving SOTA results on various datasets. 
These findings establish token-prediction based masked modeling on multi-domain datasets as a powerful framework for general audio representation learning.

\section{Acknowledgements}
This work used the Bridges2 system at PSC and Delta and DeltaAI system at NCSA through allocations CIS210014 and IRI120008P from the Advanced Cyberinfrastructure Coordination Ecosystem: Services \& Support (ACCESS) program, supported by National Science Foundation grants \#2138259,\#:2138286, \#:2138307, \#:2137603, and \#:2138296.

% \clearpage
\bibliographystyle{IEEEtran}
\bibliography{mybib}

@inproceedings{beats,
  title={{BEATs}: audio pre-training with acoustic tokenizers},
  author={Chen, Sanyuan and Wu, Yu and Wang, Chengyi and Liu, Shujie and Tompkins, Daniel and Chen, Zhuo and Che, Wanxiang and Yu, Xiangzhan and Wei, Furu},
  booktitle={ICML},
  year={2023}
}

@inproceedings{ssast,
  title={{SSAST}: Self-supervised audio spectrogram transformer},
  author={Gong, Yuan and Lai, Cheng-I and Chung, Yu-An and Glass, James},
  booktitle={AAAI},
  year={2022}
}

@article{cornell2024dcase,
  title={{DCASE} 2024 task 4: Sound event detection with heterogeneous data and missing labels},
  author={Cornell, Samuele and Ebbers, Janek and Douwes, Constance and Mart{\'\i}n-Morat{\'o}, Irene and Harju, Manu and Mesaros, Annamaria and Serizel, Romain},
  journal={DCASE Workshop},
  year={2024}
}

@article{maeast,
  title={{MAE-AST}: Masked autoencoding audio spectrogram transformer},
  author={Baade, Alan and Peng, Puyuan and Harwath, David},
  journal={Interspeech},
  year={2022}
}

@article{audiomae,
  title={Masked autoencoders that listen},
  author={Huang, Po-Yao and Xu, Hu and Li, Juncheng and Baevski, Alexei and Auli, Michael and Galuba, Wojciech and Metze, Florian and Feichtenhofer, Christoph},
  journal={NeurIPS},
  year={2022}
}

@techreport{dcase24aac_top,
    author = "Jung, Jee-weon and Zhang, Dong and Yang, Huck C.-H. and Wu, Shih-Lun and Chan, David M. and Kong, Zhifeng and Ruifan, Deng and Yaqian, Zhou and Rafael, Valle and Watanabe, Shinji",
    title = "AUTOMATIC AUDIO CAPTIONING WITH ENCODER FUSION, MULTI-LAYER AGGREGATION, AND LARGE LANGUAGE MODEL ENRICHED SUMMARIZATION",
    institution = "DCASE2024 Challenge",
    year={2024}
}

@INPROCEEDINGS{owsm,
  author={{Peng et al}},
  booktitle={ASRU}, 
  title={Reproducing Whisper-Style Training Using An Open-Source Toolkit And Publicly Available Data}, 
  year={2023},
  }

@article{openbestrq,
  title={Open Implementation and Study of BEST-RQ for Speech Processing},
  author={Whetten, Ryan and Parcollet, Titouan and Dinarelli, Marco and Est{\`e}ve, Yannick},
  journal={ICASSP workshop on Self-supervision in Audio, Speech and Beyond},
  year={2024}
}

@inproceedings{esc50,
  title={{ESC}: Dataset for environmental sound classification},
  author={Piczak, Karol J},
  booktitle={Proceedings of ACM international conference on Multimedia},
  year={2015}
}

@inproceedings{audioset,
  title={Audio set: An ontology and human-labeled dataset for audio events},
  author={Gemmeke, Jort F and Ellis, Daniel PW and Freedman, Dylan and Jansen, Aren and Lawrence, Wade and Moore, R Channing and Plakal, Manoj and Ritter, Marvin},
  booktitle={ICASSP},
  year={2017},
}

@inproceedings{espnet,
  author={Shinji Watanabe and Takaaki Hori and Shigeki Karita and Tomoki Hayashi and Jiro Nishitoba and Yuya Unno and Nelson {Enrique Yalta Soplin} and Jahn Heymann and Matthew Wiesner and Nanxin Chen and Adithya Renduchintala and Tsubasa Ochiai},
  title={{ESPnet}: End-to-End Speech Processing Toolkit},
  year={2018},
  booktitle={Interspeech},
}

@article{pengi,
  title={Pengi: An audio language model for audio tasks},
  author={Deshmukh, Soham and Elizalde, Benjamin and Singh, Rita and Wang, Huaming},
  journal={NeurIPS},
  year={2023}
}

@inproceedings{salmonn,
  title={{SALMONN}: Towards Generic Hearing Abilities for Large Language Models},
  author={Tang, Changli and Yu, Wenyi and Sun, Guangzhi and Chen, Xianzhao and Tan, Tian and Li, Wei and Lu, Lu and Zejun, MA and Zhang, Chao},
  booktitle={ICLR},
  year={2023}
}

@article{gama,
  title={{GAMA}: A Large Audio-Language Model with Advanced Audio Understanding and Complex Reasoning Abilities},
  author={Ghosh, Sreyan and Kumar, Sonal and Seth, Ashish and Evuru, Chandra Kiran Reddy and Tyagi, Utkarsh and Sakshi, S and Nieto, Oriol and Duraiswami, Ramani and Manocha, Dinesh},
  journal={EMNLP},
  year={2024}
}

@inproceedings{audioentailment,
  title={Audio Entailment: Assessing deductive reasoning for audio understanding},
  author={Deshmukh, Soham and Han, Shuo and Bukhari, Hazim and Elizalde, Benjamin and Gamper, Hannes and Singh, Rita and Raj, Bhiksha},
  booktitle={ AAAI},
  year={2025}
}

@inproceedings{audioqa,
  title={Clotho-aqa: A crowdsourced dataset for audio question answering},
  author={Lipping, Samuel and Sudarsanam, Parthasaarathy and Drossos, Konstantinos and Virtanen, Tuomas},
  booktitle={EUSIPCO},
  year={2022},
}

@inproceedings{clotho,
  title={Clotho: An audio captioning dataset},
  author={Drossos, Konstantinos and Lipping, Samuel and Virtanen, Tuomas},
  booktitle={ICASSP},
  year={2020},
}

@inproceedings{noss,
  title={Towards Learning a Universal Non-Semantic Representation of Speech},
  author={Shor, Joel and Jansen, Aren and Lang, Oran and Tuval, Omry and de Chaumont Quitry, F{\'e}lix and Tagliasacchi, Marco and Emanuel, Dotan},
  booktitle={Proc. Interspeech},
  number={2020},
  year={2020}
}

@article{vit,
  title={An image is worth 16x16 words: Transformers for image recognition at scale},
  author={Alexey, Dosovitskiy},
  journal={ICLR},
  year={2020}
}

@article{wavcaps,
  title={Wavcaps: A chatgpt-assisted weakly-labelled audio captioning dataset for audio-language multimodal research},
  author={{Mei et al}},
  journal={IEEE/ACM Transactions on Audio, Speech, and Language Processing},
  year={2024},
  publisher={IEEE}
}

@article{hubert,
  title={{HuBERT}: Self-supervised speech representation learning by masked prediction of hidden units},
  author={Hsu, Wei-Ning and Bolte, Benjamin and Tsai, Yao-Hung Hubert and Lakhotia, Kushal and Salakhutdinov, Ruslan and Mohamed, Abdelrahman},
  journal={IEEE/ACM transactions on audio, speech, and language processing},
  volume={29},
  pages={3451--3460},
  year={2021},
  publisher={IEEE}
}

@inproceedings{cochlscene,
  title={{CochlScene}: Acquisition of acoustic scene data using crowdsourcing},
  author={Jeong, Il-Young and Park, Jeongsoo},
  booktitle={Asia-Pacific Signal and Information Processing Association Annual Summit and Conference (APSIPA ASC)},
  year={2022},
  organization={IEEE}
}

@article{epickitchen,
  title={The epic-kitchens dataset: Collection, challenges and baselines},
  author={Damen, Dima and Doughty, Hazel and others},
  journal={IEEE Transactions on Pattern Analysis and Machine Intelligence},
  volume={43},
  number={11},
  pages={4125--4141},
  year={2020},
  publisher={IEEE}
}

@inproceedings{fma,
  title={{FMA}: A Dataset For Music Analysis},
  author={Defferrard, Micha{\"e}l and Benzi, Kirell and Vandergheynst, Pierre and Bresson, Xavier},
  booktitle={International Society for Music Information Retrieval Conference},
  year={2017}
}

@inproceedings{nsynth,
  title={Neural audio synthesis of musical notes with wavenet autoencoders},
  author={Engel, Jesse and Resnick, Cinjon and Roberts, Adam and Dieleman, Sander and Norouzi, Mohammad and Eck, Douglas and Simonyan, Karen},
  booktitle={ICML},
  year={2017},
}

@InProceedings{inat,
author = {Van Horn et al.},
title = {The {INaturalist} Species Classification and Detection Dataset},
booktitle = {CVPR},
year = {2018}
}

@article{fsd50k,
  title={{FSD50K}: An Open Dataset of Human-Labeled Sound Events},
  author={Fonseca, Eduardo and Favory, Xavier and Pons, Jordi and Font, Frederic and Serra, Xavier},
  journal={IEEE/ACM Transactions on Audio, Speech, and Language Processing},
  volume={30},
  pages={829--852},
  year={2022},
  publisher={IEEE}
}

@inproceedings{beans,
  title={{BEANS}: The benchmark of animal sounds},
  author={Hagiwara, Masato and Hoffman, Benjamin and Liu, Jen-Yu and Cusimano, Maddie and Effenberger, Felix and Zacarian, Katie},
  booktitle={ICASSP},
  year={2023},
}

@INPROCEEDINGS{gtzan,
  author={Mehta, Jash and Gandhi, Deep and Thakur, Govind and Kanani, Pratik},
  booktitle={2021 5th International Conference on Computing Methodologies and Communication (ICCMC)}, 
  title={Music Genre Classification using Transfer Learning on log-based MEL Spectrogram}, 
  year={2021},
  volume={},
  number={},
  doi={10.1109/ICCMC51019.2021.9418035}}

@inproceedings{biolingual,
  title={Transferable models for bioacoustics with human language supervision},
  author={Robinson, David and Robinson, Adelaide and Akrapongpisak, Lily},
  booktitle={ICASSP},
  year={2024},
}

@inproceedings{laion-clap,
  title={Large-scale contrastive language-audio pretraining with feature fusion and keyword-to-caption augmentation},
  author={Wu, Yusong and Chen, Ke and Zhang, Tianyu and Hui, Yuchen and Berg-Kirkpatrick, Taylor and Dubnov, Shlomo},
  booktitle={ICASSP},
  year={2023},
}

@inproceedings{gpm-bt,
  title={Scaling Bioacoustic Signal Pre-training with Million Samples Via Mask-Modeling},
  author={Deng, Xuyao and Wan, Tianjiao and Xu, Kele and Gao, Tian and Qiao, Peng and Feng, Dawei and Dou, Yong},
  booktitle={ICASSP},
  year={2025},
}

@inproceedings{aves,
  title={{AVES}: Animal vocalization encoder based on self-supervision},
  author={Hagiwara, Masato},
  booktitle={ICASSP},
  year={2023},
}

@article{dasheng,
  title={Scaling up masked audio encoder learning for general audio classification},
  author={Dinkel, Heinrich and Yan, Zhiyong and Wang, Yongqing and Zhang, Junbo and Wang, Yujun and Wang, Bin},
  journal={Interspeech},
  year={2024}
}

@article{mellow,
      title={Mellow: a small audio language model for reasoning}, 
      author={Soham Deshmukh and Satvik Dixit and Rita Singh and Bhiksha Raj},
      year={2025},
      journal={arXiv preprint arXiv:2503.08540},
}

@article{icme,
  title={The {ICME} 2025 Audio Encoder Capability Challenge},
  author={Zhang, Junbo and Dinkel, Heinrich and Song, Qiong and Wang, Helen and Niu, Yadong and Cheng, Si and Xin, Xiaofeng and Li, Ke and Wang, Wenwu and Wang, Yujun and others},
  journal={arXiv preprint arXiv:2501.15302},
  year={2025}
}

@inproceedings{baevski2022data2vec,
  title={Data2vec: A general framework for self-supervised learning in speech, vision and language},
  author={Baevski, Alexei and Hsu, Wei-Ning and Xu, Qiantong and Babu, Arun and Gu, Jiatao and Auli, Michael},
  booktitle={ICML},
  year={2022},
}

@inproceedings{radford2023robust,
  title={Robust speech recognition via large-scale weak supervision},
  author={Radford, Alec and Kim, Jong Wook and et al},
  booktitle={ICML},
  year={2023},
}

@inproceedings{chen2024eat,
  title={{EAT}: self-supervised pre-training with efficient audio transformer},
  author={Chen, Wenxi and Liang, Yuzhe and Ma, Ziyang and Zheng, Zhisheng and Chen, Xie},
  booktitle={Proceedings of the International Joint Conference on Artificial Intelligence},
  year={2024}
}

@techreport{beatsbiodcase,
  title={Few-shot bioacoustic event detection using beats},
  author={Gelderblom, Femke and Cretois, Benjamin and Johnsen, Pal and Remonato, Filippo and Reinen, Tor Arne},
  year={2023},
  institution={DCASE2023 Challenge}
}

@techreport{wu2023_t6a,
    Author = "Wu, Shih-Lun and Chang, Xuankai and Wichern, Gordon and Jung, Jee-weon and Germain, François and Roux, Jonathan Le and Watanabe, Shinji",
    title = "{BEATs}-based audio captioning model with INSTRUCTOR embedding supervision and ChatGPT mix-up",
    year={2023},
    institution = "DCASE2023 Challenge",
}

@inproceedings{li2024mert,
  title={MERT: Acoustic Music Understanding Model with Large-Scale Self-supervised Training},
  author={Li, Yizhi et al.},
  booktitle={ICLR},
  year={2024}
}

@article{wavlm,
  title={Wavlm: Large-scale self-supervised pre-training for full stack speech processing},
  author={Chen, Sanyuan et al.},
  journal={IEEE Journal of Selected Topics in Signal Processing},
  volume={16},
  number={6},
  pages={1505--1518},
  year={2022},
  publisher={IEEE}
}

@article{usm,
  title={{Google USM}: Scaling automatic speech recognition beyond 100 languages},
  author={Zhang, Yu and others},
  journal={arXiv preprint arXiv:2303.01037},
  year={2023}
}

@inproceedings{xeus,
  title={Towards Robust Speech Representation Learning for Thousands of Languages},
  author={Chen, William and Zhang, Wangyou and Peng, Yifan and Li, Xinjian and Tian, Jinchuan and Shi, Jiatong and Chang, Xuankai and Maiti, Soumi and Livescu, Karen and Watanabe, Shinji},
  booktitle={EMNLP},
  year={2024}
}

@inproceedings{bestrq,
  title={Self-supervised learning with random-projection quantizer for speech recognition},
  author={Chiu, Chung-Cheng and Qin, James and Zhang, Yu and Yu, Jiahui and Wu, Yonghui},
  booktitle={ICML},
  year={2022},
}

@inproceedings{vecquant,
 author = {van den Oord, Aaron and Vinyals, Oriol and kavukcuoglu, koray},
 booktitle = {Advances in Neural Information Processing Systems},
 title = {Neural Discrete Representation Learning},
 volume = {30},
 year = {2017}
}

@inproceedings{desed,
  title={Sound event detection in domestic environments with weakly labeled data and soundscape synthesis},
  author={Turpault, Nicolas and Serizel, Romain and Shah, Ankit Parag and Salamon, Justin},
  booktitle={DCASE Workshop},
  year={2019}
}

@inproceedings{us8k,
  title={A dataset and taxonomy for urban sound research},
  author={Salamon, Justin and Jacoby, Christopher and Bello, Juan Pablo},
  booktitle={Proceedings of ACM international conference on Multimedia},
  year={2014}
}

@inproceedings{fsdkgl,
  title={GENERAL-PURPOSE TAGGING OF FREESOUND AUDIO WITH AUDIOSET LABELS: TASK DESCRIPTION, DATASET, AND BASELINE},
  author={Fonseca, Eduardo and Plakal, Manoj and Font, Frederic and Ellis, Daniel PW and Favory, Xavier and Pons, Jordi and Serra, Xavier},
    booktitle={DCASE Workshop},
      year={2023}
}

@inproceedings{cider,
  title={{CIDEr}: Consensus-based image description evaluation},
  author={Vedantam, Ramakrishna and Lawrence Zitnick, C and Parikh, Devi},
  booktitle={CVPR},
  pages={4566--4575},
  year={2015}
}

@article{wav2vec,
  title={wav2vec: Unsupervised pre-training for speech recognition},
  author={Schneider, Steffen and Baevski, Alexei and Collobert, Ronan and Auli, Michael},
  journal={arXiv preprint arXiv:1904.05862},
  year={2019}
}

@article{gtzan_flaw,
  title={The {GTZAN} dataset: Its contents, its faults, their effects on evaluation, and its future use},
  author={Sturm, Bob L},
  journal={arXiv preprint arXiv:1306.1461},
  year={2013}
}

@article{versa,
  title={{VERSA}: A Versatile Evaluation Toolkit for Speech, Audio, and Music},
  author={{Shi et al}},
  journal={arXiv preprint arXiv:2412.17667},
  year={2024}
}

\end{document}